\documentclass[manuscript]{emulateapj}
\usepackage{textcomp}
\usepackage{amssymb}
\usepackage[normalem]{ulem}
\usepackage{amsmath}
\usepackage{graphicx}

\shorttitle{Dark matter free dwarf galaxies formed in jellyfish tentacles}
\shortauthors{Lora et al.}

\begin{document}

\title[Dark matter free dwarf galaxy formation\\
at the the tips of the tentacles of jellyfish galaxies.]{Dark matter free dwarf galaxy formation\\
at the the tips of the tentacles of jellyfish galaxies.}

\author{V. Lora\altaffilmark{1}, R. Smith\altaffilmark{2}, J. Fritz\altaffilmark{3}, A. Pasquali\altaffilmark{4} \& A. C. Raga\altaffilmark{1}}

\affil{\altaffilmark{1}Instituto de Ciencias Nucleares (UNAM), 
Ap. 70-264, C.P. 04510, Mexico City, Mexico}    
\affil{\altaffilmark{2}Departamento de Física, Universidad Técnica Federico Santa María,\\
Av. Vicuña Mackenna 3939, San Joaquín, Santiago, 
Chile}

\affil{\altaffilmark{3}Instituto de Radioastronom\'ia y Astrof\'isica (UNAM),
Antigua Carretera a P\'atzcuaro 8701, 58089 Morelia, Mexico}

\affil{\altaffilmark{4}Astronomisches Rechen-Institut (ZAH),
M\"onchhofstr. 12-14, Heidelberg, 69120, Germany}

\email{v.lora@nucleares.unam.mx}

\begin{abstract}{
When falling into a galaxy cluster, galaxies experience a loss of gas due to ram pressure stripping. In particular, disk galaxies lose gas from their disks and very large tentacles of gas can be formed. Because of the morphology of these stripped galaxies they have been referred to as Jellyfish galaxies. It has been found that star formation is triggered not only in the disk, but also in the tentacles of such Jellyfish galaxies. The observed star forming regions located in the tentacles of those galaxies have been found to be as massive as $3\times10^7$ M$_{\odot}$ and with sizes $> 100$ pc. Interestingly, these parameters in mass and size agree with those of dwarf galaxies. In this work we make use of the state of the art magneto-hydrodynamical cosmological simulation Illustris TNG-50, to study massive jellyfish galaxies with long tentacles. 
We find that, in the tentacles of TNG-50 Jellyfish galaxies, the star formation regions (gas+stars) formed could be as massive as $\sim2\times10^8$ M$_{\odot}$. A particular star forming region was analyzed. This region has a star formation rate of $0.04$ M$_{\odot}$/yr, it is metal rich, has an average age of $0.46$ Gyr, and has a half mass radius of $\sim1$ kpc, typical of standard dwarf galaxies. Most importantly, this region is gravitationally self-bound. 
All and all, we identify a new type of dwarf galaxy being born from the gas tentacles of jellyfish galaxies, that by construction lacks a dark matter (hereafter DM) halo.}
\end{abstract}


\section{Introduction}
Among the mechanisms that affect the evolution of galaxies as they enter galaxy clusters, ram pressure stripping (RPS, Gunn \& Got, 1972) is believed to play an important role in introducing the gas content and quenching the star formation of galaxies \citep{corteese:19}.

As a galaxy interacts
with the intra-cluster medium (ICM), the latter exerts a pressure force on the galaxy's interstellar medium (ISM), stripping it away and sometimes removing it completely. It is close to the cluster center, where the density of the ICM is the highest, and the velocity of an in-falling galaxy reaches its maximum, that RP is the most strong.

The net result of this mechanism, is the loss of a galaxy's ISM that, in turn, will form trails of gas in the direction opposite to the galaxy's motion. These have been observed at multiple wavelengths, tracing molecular and atomic gas (e.g. \citealt{jachym:14,jachym:17,jachym:19,moretti:18,moretti:20}), and ionized gas as well, appearing as H$\alpha$ emission (e.g. \citealt{fossati:16,mcpartland:16,poggianti:19}). Clumpy H$\alpha$ can indicate the presence of active star formation outside the galaxy. The presence of young, massive stars in the tails, has also been traced in the UV and optical bands (e.g. \citealt{kenney:14,george:18,poggianti:19}). The morphology these galaxies present while undergoing stripping, led them to be dubbed "Jellyfish galaxies".

Long, one sided tails, that often tend to point away from the cluster center, is one of the tell-tale indicators of ram pressure.
As RPS is a purely an hydrodynamical force, only the gas component is directly influenced by it, while the stars will go largely unaffected both, from a geometrical and from a dynamical point of view. By this token, the detection of a recently formed young stellar component in the tails of these objects, strongly suggests that this SF occurs in-situ within the stripped gas.

The GASP survey (GAs Stripping Phenomena in galaxies with  MUSE; \citealt{poggianti:17}) has provided us with one of the most detailed views of the properties of such unique type of galaxy 
in the local ($0.04<z<0.07$) universe. Among the $\sim 60$ RPS galaxies belonging to this sample, ionized gas powered by star formation has been detected 
(R: you can get ionised gas that is not a result of SF as well, so maybe you could sya how they have provided detailed information about the properties of the ionised gas, its dynamics and SF in the tails instead) in the tentacles, when the latter are present
\citep{gullie:20}. Furthermore, in a few cases, the superb resolution provided by ALMA has allowed the detection of CO within the same location \citep{moretti:18,moretti:20}, hence further supporting the idea that stars can be formed in this environment, outside of the galaxy. 

Among the works analyzing this issue in the framework of GASP, \cite{gullie:20} studied the star formation in the tentacles, and found an average star formation rate (hereafter SFR) value of $0.22$ M$_{\odot}$/yr per galaxy cluster. In addition, they estimated a total mass of stars formed in the tentacles of $4\times10^9$ M$_{\odot}$ per galaxy cluster since $z\sim 1$.

\begin{figure*}
    \center
    {\includegraphics[width=0.67\textwidth]{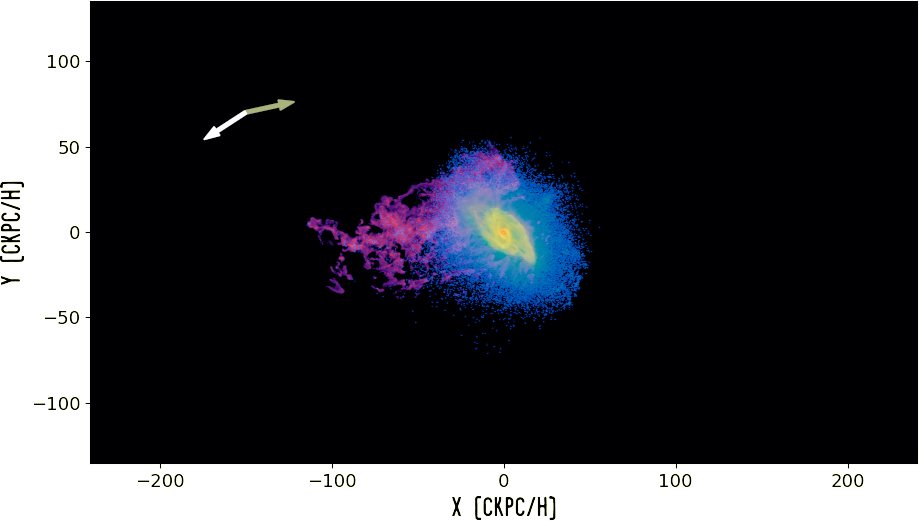}}
    {\includegraphics[width=0.159\textwidth]{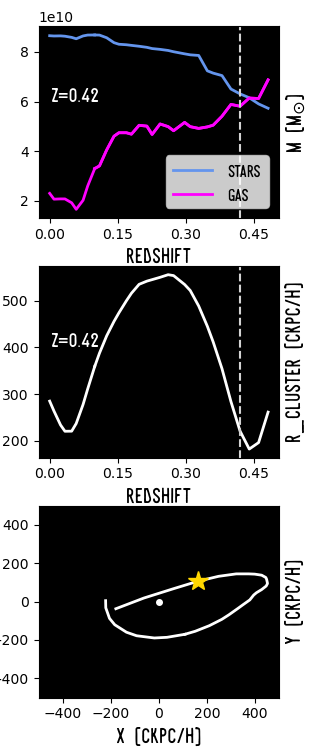}}\\
    {\includegraphics[width=0.67\textwidth]{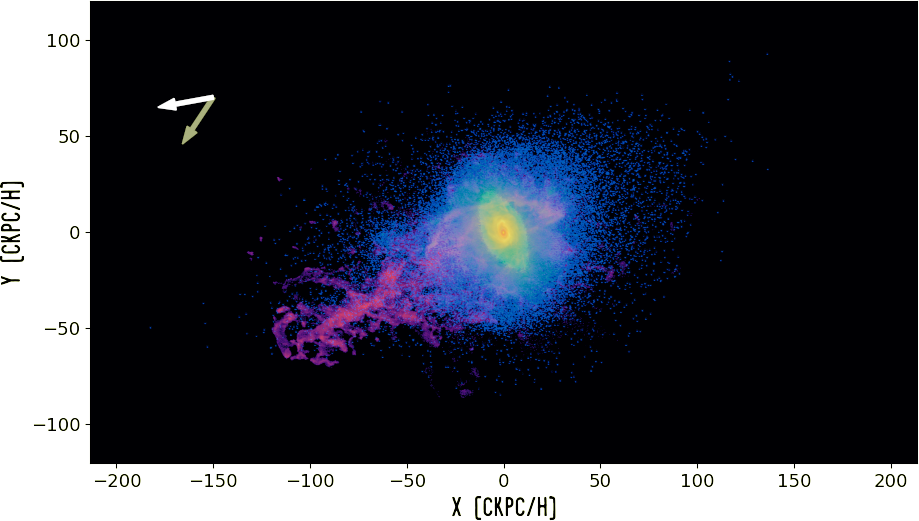}}
    {\includegraphics[width=0.159\textwidth]{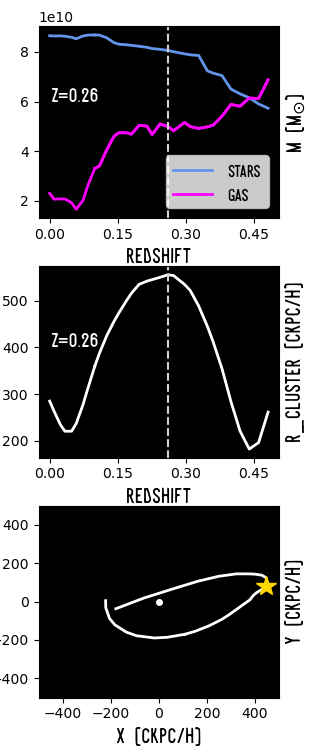}}\\
    {\includegraphics[width=0.67\textwidth]{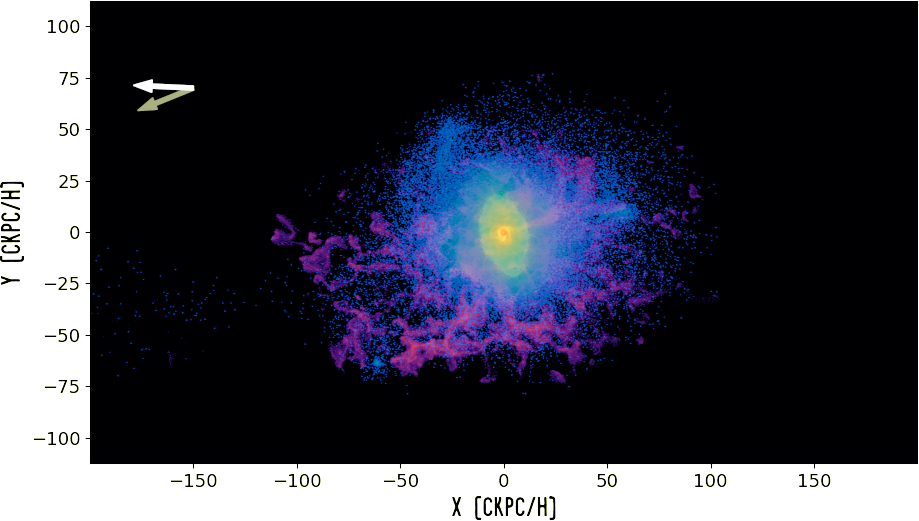}}
    {\includegraphics[width=0.159\textwidth]{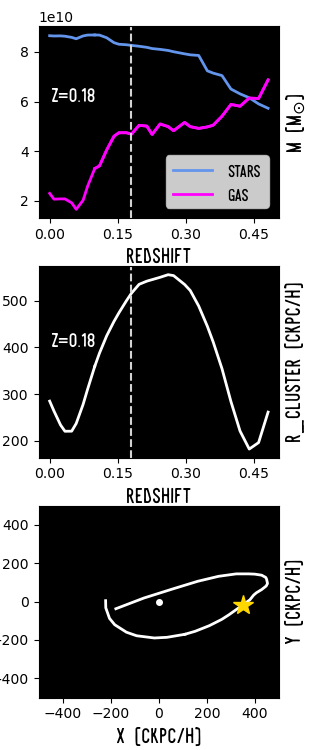}}
    \caption{Maps of the stellar (yellow-blue), and gas (pink-purple) distributions at the evolutionary times z=0.42 (top panel), z=0.26 (middle panel), and z=0.18 (bottom panel). The white arrows indicate the direction to the center of the cluster of galaxies. The gray arrows indicate the direction of motion of the galaxy with ID 119447. The corresponding evolutionary time (given in redshift) is shown in the lower left corner of each panel. 
    The three sub-panels located at the right side of The Figure: (top sub-panel) the gas (magenta) and stellar (blue) mass as a function of time (redshift). The middle sub-panel shows the cluster-centric distance of the galaxy as a function of time (redshift). The bottom sub-panel shows the $X$-$Y$ orbit of the galaxy around the galaxy cluster center. The small white dot shows the center of the cluster, and the yellow star shows the position on the orbit of the galaxy, at the time where the galaxy maps are plotted.}
    \label{Fig1}
\end{figure*}  

Going into further details, \cite{poggianti:19} studied star formation in the tails, by focusing on star forming clumps, and found that their stellar masses range from $10^5$ to $3\times10^7$ M$_{\odot}$ (with a median of $3\times10^6$ M$_{\odot}$), and that their core radius range from $100$ to $400$ pc (with a median of $160$ pc), similar in masses and sizes to ultra compact dwarf galaxies (UCDs).

Further studies from the same group, exploiting HST follow-up observations of galaxies that are extreme examples of stripping \citep{gullie:23}, studied the characteristics of the star clumps as a population, finding that star formation in this environment is turbulence--driven, something that is found to be common in main-sequence galaxies as well \citep{giunchi:23}.

The results of these observational works, prompted us to study and analyze star formation in the tentacles of jellyfish galaxies detected in Illustris TNG-50 simulation, a state of the art cosmological simulation. The main goal of this paper, is to look
for the presence of self-gravitating objects in the tails of RPS-affected galaxies, whose masses and size resemble those of standard dwarf galaxies. The discovery of such objects in an magneto-hydrodynamical simulation, may allow us to better understand the sequence of events that lead up to the formation of such objects, and further support the
hypothesis that ram pressure is a viable mechanism for the production of RPS dwarf galaxies (DG), a secondary type of DG in the same family as the so called tidal dwarf galaxies (TDGs), which are formed from tidal debris of interacting galaxies.

This RPS DGs, whose existence was already speculated in \cite{poggianti:19}, would be dark matter (DM) free by definition, and would display properties that fall in the range of those observed between UCDs and standard dwarfs.

This article is organized as follows: In Section \S\ref{sec:methods} we describe the TNG-50 simulation, and describe our selection method to identify jellyfish galaxies.
In Section \S\ref{sec:results} we present our results. Finally, we present our conclusions in Section \S\ref{sec:conclusions}.

\begin{figure*}
    \center
    {\includegraphics[width=0.67\textwidth]{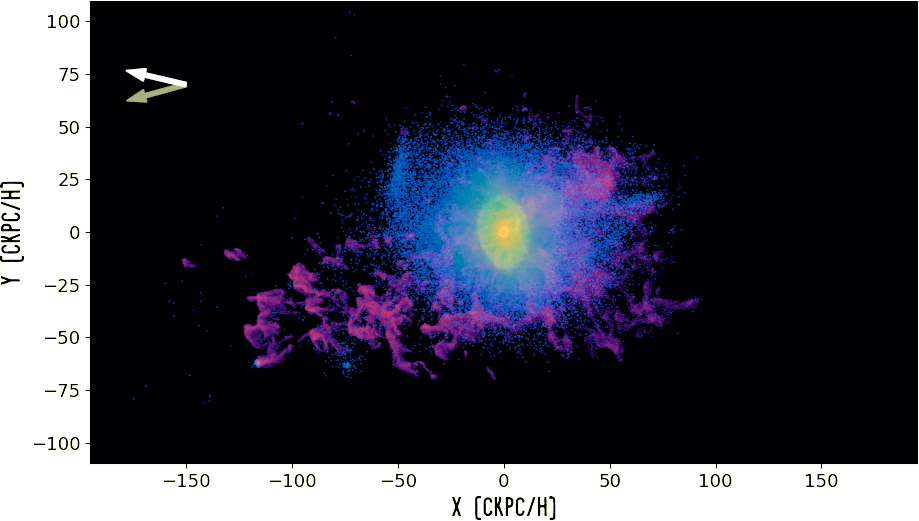}}
    {\includegraphics[width=0.159\textwidth]{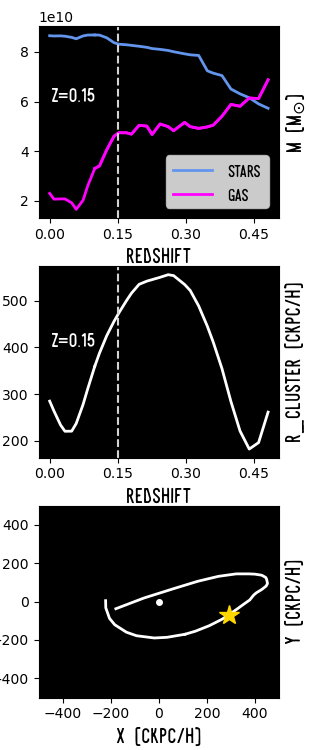}}\\
    {\includegraphics[width=0.67\textwidth]{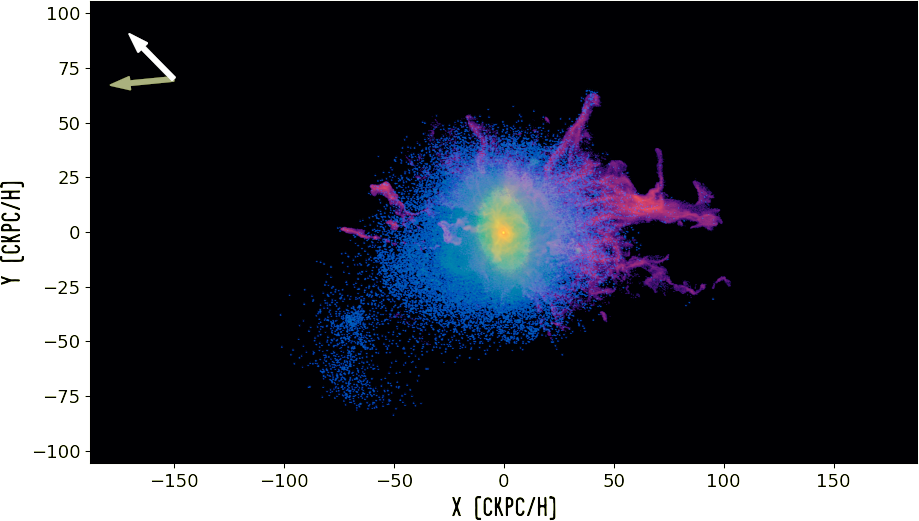}}
    {\includegraphics[width=0.159\textwidth]{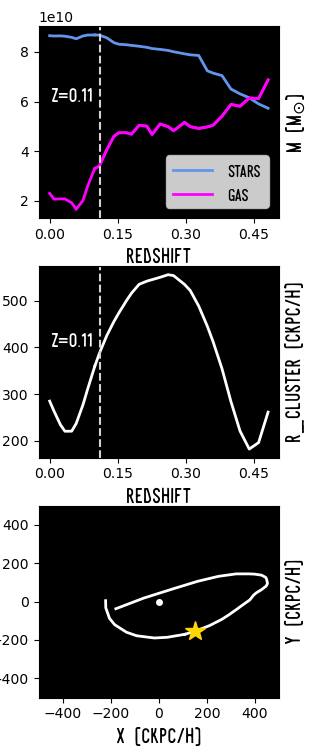}}\\
    {\includegraphics[width=0.67\textwidth]{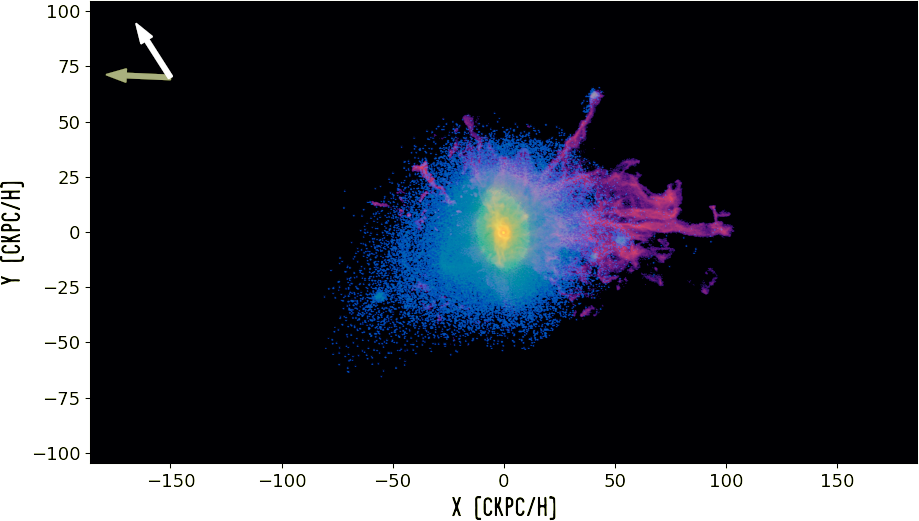}}
    {\includegraphics[width=0.159\textwidth]{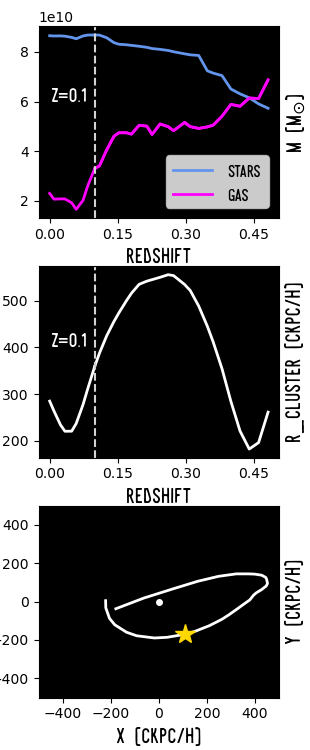}}
    
    \caption{Same as Figure \ref{Fig1}, but at the evolutionary times z=0.15 (top panel), z=0.11 (middle panel), and z=0.1 (bottom panel).}
    \label{Fig2}
\end{figure*}

\section{The TNG-50 simulation and the sample selection}
\label{sec:methods}
In this work we study the regions of star formation in the stripped gas from jellyfish galaxies at a cosmological time of z=0.1 in TNG-50 \citep{pillepich:19,nelson:19}. In the following sub-sections we briefly describe the TNG-50 simulation, and our sample selection.

\subsection{Illustris TNG-50}
The Illustris TNG-50 is the simulation with the highest resolution in the  Illustris-TNG suite of cosmological-magneto-hydrodynamical simulations \citep{marinacci:18,naiman:18,nelson:18,pillepich:18,springel:18}. The TNG simulations are run with the Arepo MHD adaptive code \citep{springel:10}. The simulations use a $\Lambda$-CDM frame work, with $\Omega_m=0.3089$, $\Omega_b=0.0486$, $\Omega_{\Lambda}=0.6911$,  $h=0.6774$, $\sigma_{8}=0.8159$, and $n_s=0.9667$, which are the matter density, the baryonic density, the cosmological constant, the Hubble constant, the normalization, and the spectral index, respectively, as given by the Planck data \citep{planck:16}.

The TNG simulations consist of a set of simulations with different domain sizes, and resolutions. In particular TNG-50 has a ($51.7$ Mpc)$^{3}$ volume box, with a baryonic mass resolution $m_b = 8.5\times10^4$ M$_{\odot}$, and a DM mass resolution $m_{DM} = 4.5\times10^5$ M$_{\odot}$. The minimum allowed adaptive gravitational softening length for gas cells (co-moving Plummer equivalent) is $\epsilon_{gas,min}= 74$ pc and for the stars and DM  $\epsilon_{DM,*}= 288$ pc.

The most important physical ingredients included in the TNG simulations are: The gas radiative processes, the star formation in the dense interstellar medium, the evolution of the stellar population and the chemical enrichment from supernovae Ia, II, as well as from AGB stars \citep{nelson:19}.

However, because of the resolution limit in TNG-50 there are physical processes like small-scale turbulence, thermal instabilities, and molecular cloud formation, which can not be explicitly modelled \citep{vogelsberger:13}.

The star formation is modelled with a density threshold ($0.1$ cm$^{-3}$) as described in \cite{springel:03}. In such star formation recipe the gas parcels are stochastically converted into star particles when their density is greater than $n_{H}=0.1$ cm$^{-3}$ \citep{kennicutt:83} on a time-scale proportional to the local dynamical time of the gas.

Jellyfish galaxies have been studied before in the TNG-50 context. For example, \cite{rohr:23} analyzed a set of first-infalling jellyfish galaxies in TNG-50, where they find that jellyfish galaxies are a significant source of cold gas accretion into the ICM. Moreover, \cite{goller:23} confirmed that star formation can be triggered within the RPS tentacles of jellyfish galaxies despite of not finding an overall larger SFR in jellyfish galaxies, compared to their control galaxy sample.

It is true that in some regions the media of the RPS tails could be more diffuse as compared to that of the inter-stellar medium ($\sim$1 particle/cm$^3$). 
However, in our case the gas density in some regions of the tentacles of the main galaxy can be dense enough to reach the threshold imposed by TNG-50 to form stars. In this case then, the sub-grid recipes used in TNG-50 do not affect our results.

In summary, the TNG-50 simulation combines a large-scale volume with a high mass resolution, a powerful tool to study large star formation regions in the tails of jellyfish galaxies.




\subsection{Sample selection}
\label{subsecsec:sample}
In this study the jellyfish galaxies are identified at a redshift z=0.1. We follow the criteria in \cite{yun:19}, which select satellite galaxies belonging to massive galaxy clusters, (hereafter FoF groups; Friends of Friends groups) with a total mass $>10^{13}$ M$_{\odot}$. They consider only those galaxies whose positions are located between $>0.25$ R$_{200}$ and $<$ R$_{200}$ from the center of the galaxy cluster (R$_{200}$ denotes the virial radius of the FoF group). The mass of the stellar component of the jellyfish candidate is selected to be $> 10^{9.5}$ M$_{\odot}$, so that a minimum number of stellar particles (3 thousand) is assured, with the resolution of TNG-100. We consider only the galaxies whose total gas mass to total star mass ratio ($M_{gas}/M_{stars}$) is greater than $0.01$. Additionally, \cite{yun:19} filter their sample by visual inspection requiring an asymmetric distribution of galaxy gas elongated in a preferred direction, and require no companion (interacting) galaxy to avoid tidal effects.  

At $z=0.1$, there are $23$ FoF groups that meet the jellyfish group mass criteria stated above. Applying the rest of the criteria, one ends up with $442$ galaxies. Since we are interested in the star formation within the tails of jellyfish galaxies, from that total group of $442$ galaxies we select only the most massive ones with total masses greater than $2.5\times10^{11}/h$ M$_{\odot}$, since we want the most massive star formation regions ($>10^7$ M$_{\odot}$). We end up with $23$ galaxies, which have an average gas mass of $2\times10^{10}/h$ M$_{\odot}$, an average stellar mass of $9\times10^{10}/h$ M$_{\odot}$, an average DM mass of $5.3\times10^{11}/h$ M$_{\odot}$, and an average total mass of $6.4\times10^{11}/h$ M$_{\odot}$.

From the $23$ galaxy sample, we find that seven galaxies have a galaxy neighbour (or multiple neighbours) within $50$ ckpc from their centers. We eliminate those galaxies, since we want to isolate the effect of RPS triggering large regions of star formation. The latter reduces the sample to $15$ galaxies.

In order to identify regions of star formation within the tentacles of jellyfish galaxies (massive enough to form dwarf galaxies), one must select those galaxies which have larger amounts of total gas outside their disks. 
Typically the jellyfish stage occurs at the moment when the RPS effect on a galaxy is maximum (i.e. when the galaxy falls into the cluster of galaxies). Since stars do not undergo the RPS effect, a measure of jellyfish-ness could be the ratio of the half-mass radius of the stellar component to the half-mass radius of the gas component. We computed the stellar to total gas half-mass ratio for all $15$ galaxies from the sample and tagged as candidates those galaxies whose stellar to gas half-mass ratio $r_{h,*}/r_{h,gas} < 1/3$ (the gas distribution of the galaxy is extended beyond the disk).

We found six galaxies which have gas half-mass radius three times larger than their stellar counterparts. But, only three of them present an extended distribution of gas. Those extended (tentacle-like) regions of gas are our targets to look for large regions of star formation outside the host galaxy. The further away the star forming regions in the tentacles of a jellyfish galaxy are from the host galaxy, the better chance star formation regions have to survive (tidal) gravitational interactions with their host galaxy. From those three galaxies with extended gas distribution,  we looked for stellar and gas over-densities within their jellyfish tentacles and found evidence for their presence in one of those three galaxies. This galaxy is the object of our study.

\begin{figure*}
    \center
    {\includegraphics[width=0.99\textwidth]{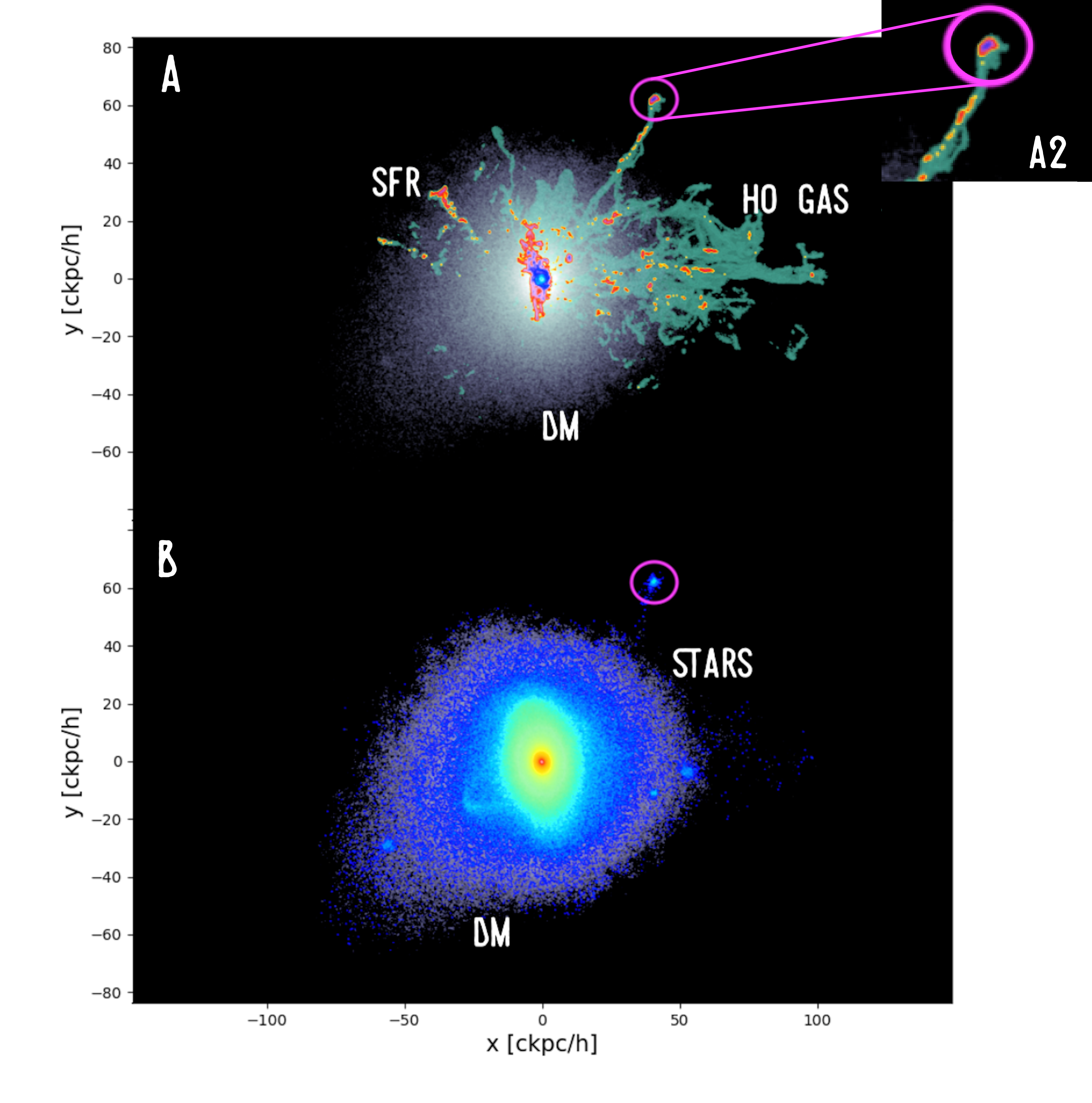} }
    \caption{The top panel (A) shows overlapping maps of neutral gas (green), SFR (rainbow), and DM (white) of the galaxy with ID 119447 at z=0.1. In magenta we circle the ram pressure stripped dwarf galaxy candidate. Panel A2 shows a close-up of the location of the ram pressure stripped dwarf galaxy.
    The bottom panel (B) shows the overlapping maps of stellar mass (rainbow) and DM (white) of the galaxy with ID 119447. Again, the magenta circles the ram pressure stripped dwarf galaxy candidate.}
    \label{Fig3}
\end{figure*}

\section{Results}
\label{sec:results}
We analysed the galaxy with ID 119447, that full-fills the criteria presented in Section \ref{sec:methods}. 
This galaxy has a total mass of $4.16\times10^{11}$ M$_{\odot}$ (with M$_{DM}=3\times10^{11}$ M$_{\odot}$, M$_{gas}=2.3\times10^{10}$ M$_{\odot}$, and M$_{*}=8.65\times10^{10}$ M$_{\odot}$). It has a $SFR = 2.14$ M$_{\odot}$/yr, and a half-mass radius of $19$ kpc.

In Figures \ref{Fig1} and \ref{Fig2}, we show maps of the stellar (yellow-blue), and gas (pink-purple) distributions at six different evolutionary times (from top to bottom redshifts z=0.42, 0.26, and 0.18; and z=0.15, 0.12, and 0.1, respectively). The white arrows indicate the direction to the center of the cluster of galaxies, while the gray arrow indicates the direction of motion of the galaxy with ID 119447. Each map in Figures \ref{Fig1} and \ref{Fig2} come with three panels. The top panel shows the amount of gas (magenta), and stars (blue) as a function of time (given in redshift). The gray vertical line shows the redshift at which the galaxy maps are plotted. The middle panel shows the cluster-centric distance of the galaxy, as a function of time (given in redshift), the vertical gray line corresponds again to the redshift at which the galaxy maps are plotted. The bottom panel shows the $X$-$Y$ orbit of the galaxy around the galaxy cluster center. The small white dot shows the center of the cluster, and the yellow star shows the position on the orbit of the galaxy of the galaxy, at the particular time when the galaxy maps are plotted. The white arrows in all panels represent the direction to the center of the galaxy cluster, and the gray arrows represent the direction of motion of the galaxy.

The top panel of Figure \ref{Fig1} shows the galaxy when it reaches its first peri-cluster passage (infall) to the galaxy cluster, showing a slightly increase of stellar mass and a decrease in gas mass. We observe that the gas distribution is quite deformed due to RPS already at this time (z=0.42). The middle panel in  Figure \ref{Fig1} shows the galaxy at its apo-cluster distance (z=0.26), where the galaxy changes direction (gray arrow) and starts to move towards the center of the galaxy cluster. Again the gas distribution appears quite disturbed. Figure \ref{Fig2} shows three snapshots of the galaxy moving in its orbit as it approaches a second infall. Already at z=0.15, the gas of the galaxy starts to be stripped (see pink lines in top subpanels of Figure \ref{Fig2}). Between z=0.12 and z=0.1, we observe a major decrease in gas while the stellar component slightly increases. In the latter period of time a prominent tentacle is formed.

At z=0.1 (bottom panel of Figure \ref{Fig2}), the galaxy with ID 119447 presents clear gas and stellar over-densities. In the top panel of Figure \ref{Fig3}, we show overlapped maps of neutral gas (green), SFR (rainbow), and DM (white). We observe that long ($\sim80$ ckpc/h) tentacles of neutral (green) gas emerge outside the DM distribution of the galaxy 119447. 

In Figure \ref{Fig3}, we circle in magenta a particular region at the tip of one prominent gas tentacle. Star formation is present along this tentacle. At the tip of the tentacle (see a close-up of tip region in question; A2 in the top panel of Figure \ref{Fig3}), the SFR is $\sim0.03$ M$_{\odot}$/yr, which is as large as the SFR at the outskirts of the disk of the galaxy 119447 (see dark blue-pink colors in the top panel of Figure \ref{Fig3}). We also observe that this star formation region is quite extended.

The magenta circle, at bottom panel of Figure \ref{Fig3} (B) shows the stellar mass map over-plotted to the DM mass map. The tip of the gas tentacle is present here as a stellar over-density. 

The fact that stars do not trace the tentacle, gives us a hint that the tentacle has a RPS nature, and that tidal effects (e.g. galaxy-galaxy gravitational interactions) were not involved. To make sure that in the formation of the tentacle of gas no tidal effects were involved, we computed the cluster-centric distance of our galaxy (ID 119447) from z=1.5 to z=0.1. Our galaxy experienced two infalls to the cluster. The first peri-cluster distance happens at z=0.42 (see top panel of Figure \ref{Fig1}), and the second infall happens at z=0.05. Both infalls of the galaxy into the cluster, are characterized by only gas loss. No stellar stripping happens while our galaxy falls towards the cluster centre (see top right sub-panels in each panel of Figures \ref{Fig1} and \ref{Fig2}). 

In addition in the time span analyzed and shown in Figures \ref{Fig1} and \ref{Fig2} (z=0.42-0.1), the galaxy with ID 119447 has not experienced a major merger. We conclude that the reduction in gas (but not in the stars) has a pure ram pressure nature. It is important to notice that the tentacle where we find the stellar and gas over-density lays well outside the DM halo of the galaxy (see white distribution in Figure \ref{Fig3}).

The region within the magenta circle shown in Figure \ref{Fig3}, is a very promising candidate to be a dwarf galaxy born from a RPS tentacle formed while the galaxy with ID 119447 entered its cluster of galaxies for the second time.

It has to be noted that few smaller stellar over densities appear linked to smaller gas tentacles. For example, in the top and middle panels of Figure \ref{Fig2} one can see in the bottom-left region a small stellar over-density linked to a thin and short gas tentacle. 


\begin{figure}[h!]
    \centering
    {\includegraphics[width=0.5\textwidth]{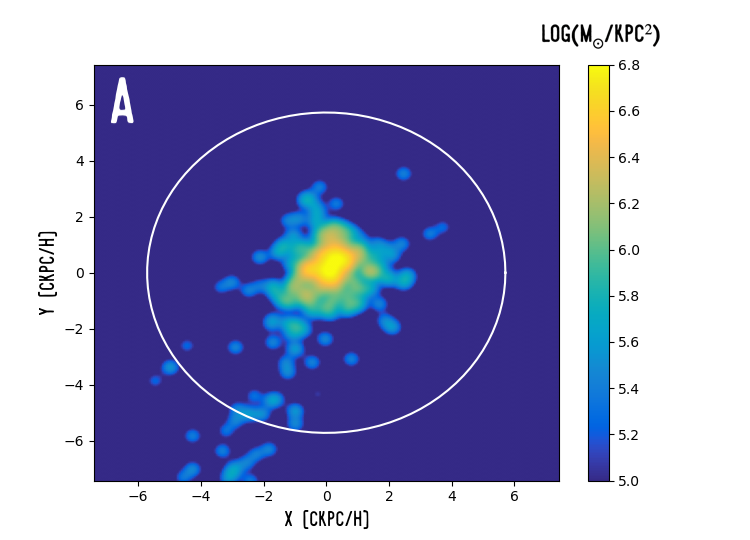}}\\
    {\includegraphics[width=0.45\textwidth]{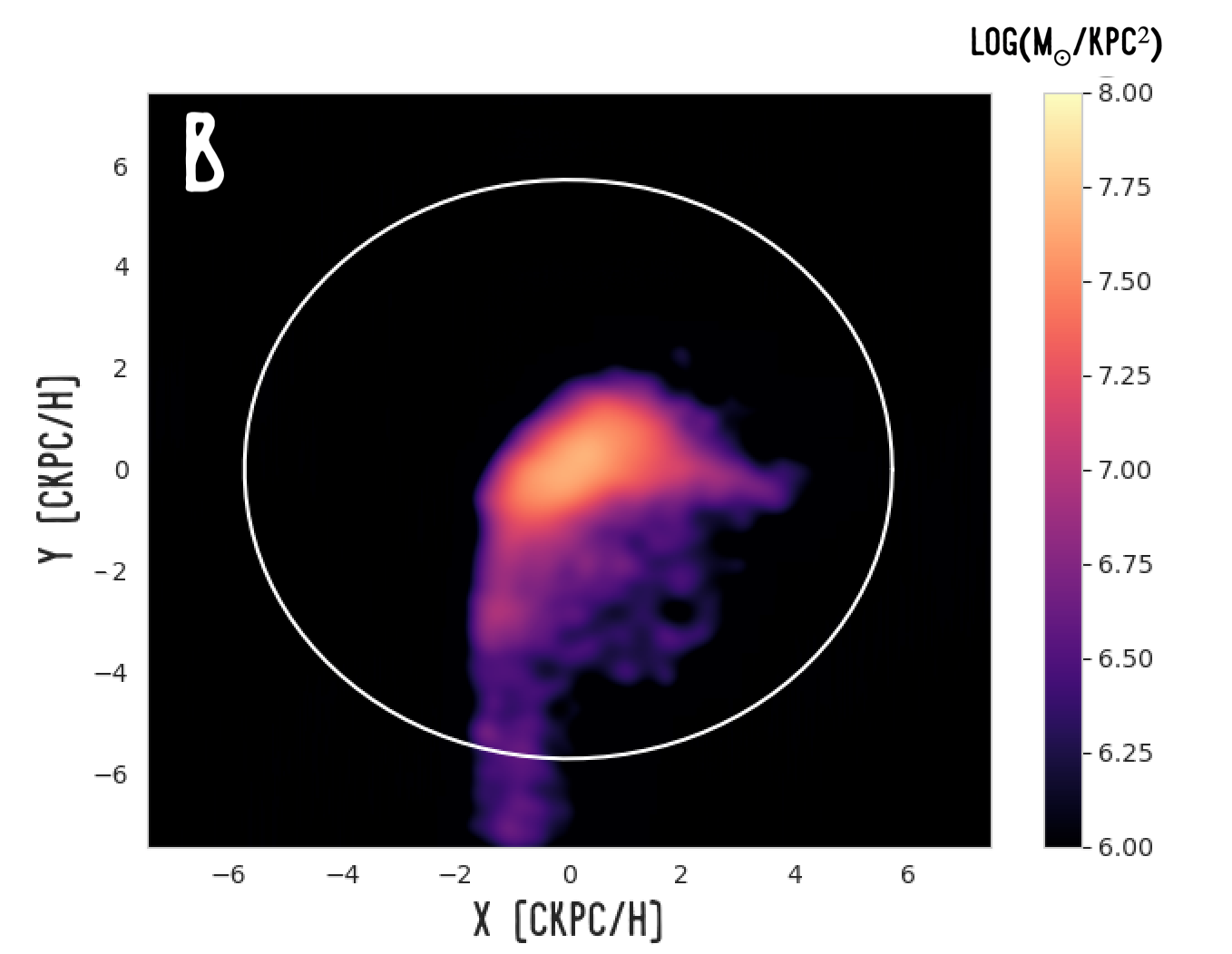}}\\
    {\includegraphics[width=0.44\textwidth]{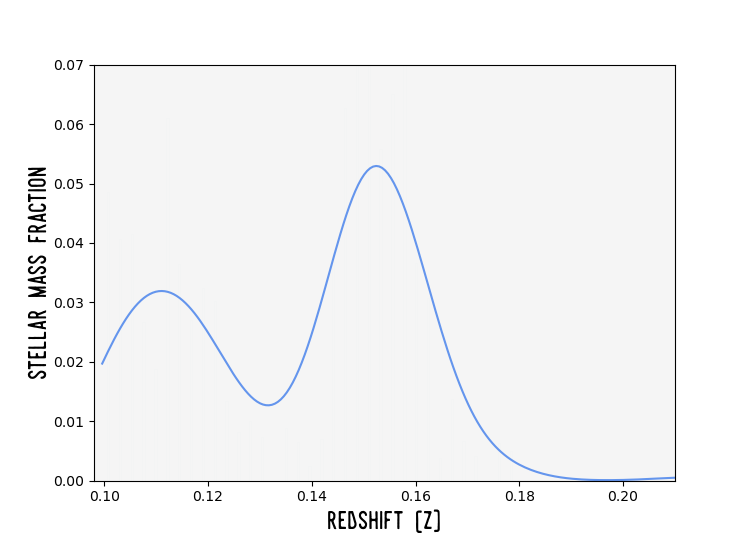}}
    \caption{The top panel (A) shows the $XY$ stellar mass map of the ram-pressure-stripped dwarf galaxy candidate. The middle panel (B) shows the $XY$ map of the gas mass. 
    In both upper panels (A and B) the white circle has a radius of $\approx 5.7$ ckpc/h ($1.5$ times the stellar-half-mass radius of the galaxy with ID 119447). The bottom panel shows the smoothed stellar mass fraction of the ram-pressure-stripped dwarf galaxy, as a function of redshift, showing two main episodes of star formation at $z\sim0.15$ and $z\sim0.11$.}
    \label{Fig4}
\end{figure}

\subsection{The Dwarf galaxy candidate}
\label{subsec2C}
We trace back the gas that forms the tentacle from where we find the stellar and gas over-density (highlighted with a white contour in Figures \ref{Fig1} and \ref{Fig2}). In the first infall of the galaxy at z=0.42 (see top panel of Figure \ref{Fig1}) the gas has been already stripped from the galaxy. Then, the stripped gas is blown across the galaxy by the change in the galaxy direction (see bottom panel of Figure \ref{Fig1}). Later, the gas in the galaxy gets stripped again (see Figure \ref{Fig2}) as it approaches the second infall (peri-center) in the opposite direction of motion. The middle and bottom panels of Figure \ref{Fig2}, show the formation of the RPS tentacle.

The formation of the tentacle and eventually the formation of the stellar over-density found at the tip of it, goes back in time since the first infall of the galaxy, where the galaxy has already had a jellyfish phase at z=0.42.

We analyze in more detail the tip of the tentacle described in the previous subsection, located at a distance of $75$ ckpc/h from the center of the galaxy with ID 119447. We built a map of stellar mass centered at the center of mass of this region, shown in the top panel (A) of Figure \ref{Fig4}. In this Figure, the white circle represents $1.5$ times the stellar-half-mass radius of the galaxy with ID 119447 ($5.7$ ckpc/h). In the bottom panel (B) of Figure \ref{Fig4}, we show the gas mass map of the SF region within the magenta circle in Figure \ref{Fig3} (labeled as A2). 

In order to check whether this SF region with an over-density in stellar mass is in fact an independent object, we have to make sure that the region is self-bound. We computed the total mass fraction of stars and gas as a function of the total energy of the substructure. We tagged the particles with total energies lower than zero, since they define a self-bound and independent object. The total stellar-self-bound mass is $M_{*,dwarf} = 1.7\times10^7$ M$_{\odot}$ (253 stellar particles). The total gas-self-bound mass is $M_{gas,dwarf} = 2\times10^8$ M$_{\odot}$ (2036 gas particles). Then, our self-bound region has a total baryonic mass of $2.17\times10^8$ M$_{\odot}$, and a gas fraction of $\sim90$\%. 

Our assumption is that self-bound objects formed at the tips of jellyfish tentacles should be DM free by construction. To prove it to be true, we computed the number of DM particles within a radius of $5.7$ ckpc/h (white circles in Figure \ref{Fig4}).  Within that volume we found $38$ DM particles ($\approx 1.7\times 10^{7}$  M$_{\odot}$). We then computed the total energy associated to these DM particles, and found that none of them were bound; i.e. $E_{tot,DM} > 0$. 


In order to study the star formation activity of the self-bound region, we built the instantaneous SFH as a function of the redshift \footnote{We smoothed the stellar mass fraction (Figure \ref{Fig4}) with a Gaussian kernel density estimation, with a smoothing bandwidth equal to 1.}. The SFH is presented in the bottom panel of Figure \ref{Fig4}. There are clearly two star formation episodes. The first one peaks at z=0.15 (see top panel of Figure \ref{Fig2}), when the gas from the main galaxy has piled up, as the galaxy moves forward from the apocenter and experiences RPS. The second episode peaks at z=0.11, when a prominent tentacle of gas starts forming (see middle panel of \ref{Fig2}). The mean (mass weighted) age of the stars in our self-bound region is $0.46$ Gyr.

The next step was to build the cumulative mass profile of the particles belonging to our self-bound region (A2 of Figure \ref{Fig3}). We computed a stellar half-mass radius  $r_{h,*} \approx 1$ ckpc/h, and a gas half-mass radius  $r_{h,gas} \approx 1.45$ ckpc/h. These sizes, together with the stellar mass and the gas mass of the self-bound region, are consistent with those of a ``standard'' dwarf galaxy. In view of these characteristics, \textit{our self bound-region could be classified as a new type of dwarf galaxy: a RPS dwarf galaxy, which by construction lacks of a DM halo.}  

\begin{figure}
    \centering
    {\includegraphics[width=0.45\textwidth]{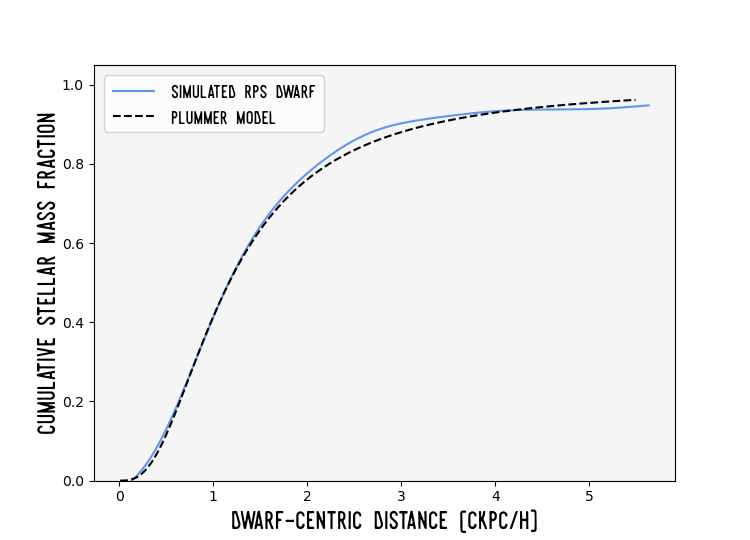} }\\
    {\includegraphics[width=0.45\textwidth]{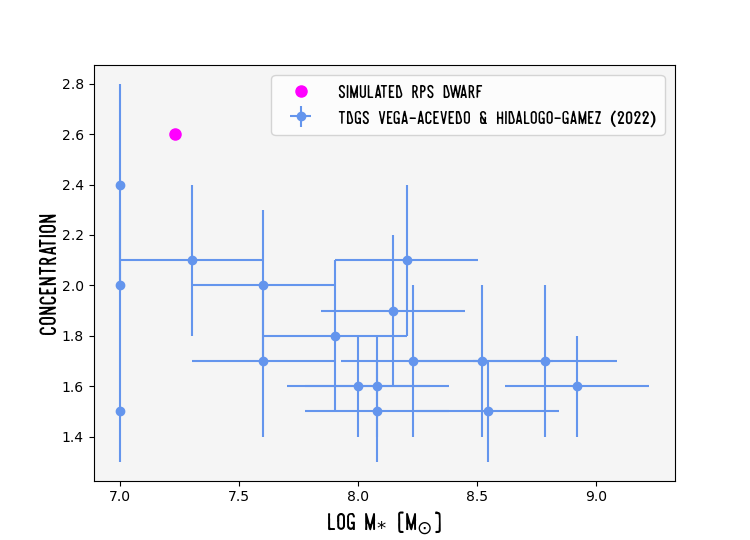}}\\
    {\includegraphics[width=0.45\textwidth]{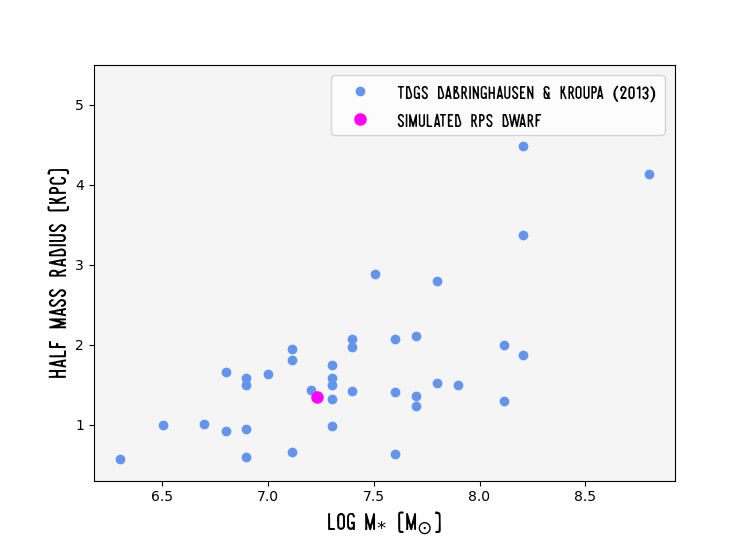}}
    \caption{The top panel shows the cumulative stellar mass fraction of the ram-pressure-stripped dwarf galaxy (blue line). The black dashed line is the analytical Plummer model with a scale factor $a=0.89$ ckpc/h (see text). 
    The middle panel shows the concentration parameter as a function of the total stellar mass, for the sample of TDGs reported in Vega-Acevedo \& Hidalogo-Gamez (2022) (blue points). The concentration parameter for our ram-pressure-stripped dwarf galaxy is shown as a magenta point.
    The bottom panel shows the half-mass radius as a function of stellar mass in the TDGs sample reported in Dabringhausen \& Kroupa (2013) (blue points). The magenta point shows the values of our ram-pressure-stripped dwarf galaxy.}

    \label{Fig5}
\end{figure}

As a next step to a better characterization of our RPS-dwarf candidate, we computed its cumulative mass profile, which we show as a blue line in Figure \ref{Fig5}. We compare the mass distribution obtained, with the mass distribution of a Plummer sphere. The Plummer mass distribution relates the half-mass radius with the scale length (r$_h$=1.3$\times$a), therefore for our RPS-dwarf galaxy, $a=0.89$ ckpc/h. We show the analytical Plummer mass distribution as a black dashed line in Figure \ref{Fig5}. Our RPS-dwarf galaxy presents a mass distribution that can be approximated by a Plummer mass distribution, as spherical systems (i.e. elliptical dwarf galaxies, spheroidal dwarf galaxies,  globular clusters, etc.) do.

A pertinent comparison of the properties of our RPS-dwarf galaxy, would be with those of TDGs, since TDGs also form without the need of a DM halo, and their gas (and consequently their in-situ stars) is already enriched.

Therefore, we compare the concentration of our RPS-dwarf galaxy to the concentration of TDGs reported by \cite{vega:22}. The concentration parameter is defined as $C=5\times log(r(80\%)/r(20\%))$, where r(80\%) is the radius at which 80\% of the stellar mass is contained. Analogously, r(20\%) is the radius at which 20\% of the stellar mass is contained. For our RPS-dwarf galaxy we computed $C = 5 \times log10(2.122/0.632) = 2.63$. In Figure \ref{Fig5} we compare the concentration value of our RPS-dwarf galaxy, with those of the TDGs reported in \cite{vega:22}. Even if the concentration value is high for our RPS-dwarf galaxy (magenta circle in Figure \ref{Fig5}), it still has a very similar value to the TDG Arp305E  ($C=2.4 \pm 0.4$, and M$_{*}=10^7$ M$_{\odot}$) reported in \cite{vega:22}.

As we mentioned before, the half-mass radius of the stellar component of our RPS-dwarf galaxy is $\sim 1$ ckpc/h. Then, we compare it with the TDGs sample of \cite{dabringhausen:13}. In the bottom panel of Figure \ref{Fig5}, we show the half-mass-radius as a function of the stellar mass. The magenta point represents the value for our RPS-dwarf galaxy, and the blue points show the corresponding values of the TDG's sample reported in \cite{dabringhausen:13}.

Regarding the rotation, TDGs can show strong velocity gradients which are consistent with rotation \citep{weilbacher:02,bournaud:04,bournaud:07}. However, it has to be noted that the TDGs for which rotation was reported, have a total mass at least an order of magnitude larger than our RPS dwarf galaxy. To compute the rotation profile of our RPS dwarf galaxy, we first identify the position angle (PA) of the rotation axis in the plane of the sky of our RPS dwarf galaxy. We followed the approach of \cite{cote:95} (see also, \cite{bellazzini:12} and \cite{bianchini:13}). We computed  a $PA=50^{\circ}$ and an amplitude (A) of $2.6$ km/s, which gives an estimate of the internal rotation $V_r$ \citep{bianchini:13}. Additionally, we computed the value of the velocity dispersion $\sigma=10.6$ km/s for our RPS dwarf galaxy. The velocity dispersion is very similar to those simulated TDGs in \cite{ploeckinger:15} which range from $\sim15$ to $\sim6$ km/s. Now, we can compare the ordered and the random motions in our RPS dwarf galaxy by computing the ratio $V_r/\sigma=0.25$, a value similar to those of large globular clusters (e.g. $M15$ has a $V_r/\sigma=0.23$ \cite{bianchini:13}). A system whose gravitational support is rotation-dominated has $V_r/\sigma>1$, for example $V_r/\sigma =1-2$ for the TDGs reported by \cite{lelli:15}.



We proceed to compute the total SFR of our RPS-dwarf galaxy, obtaining a mean SFR of $0.04$ M$_{\odot}$/yr. Using the panchromatic STARBurst IRregular Dwarf Survey \citep[STARBIRDS;][]{mcquinn:15A} \cite{mcquinn:15B} reported the SFRs and stellar masses of starburst and post-starburst dwarf galaxies. They compare the stellar mass of their starburst dwarf galaxies as a function of their SFR. Comparing the values of SFR and stellar mass of our RPS-dwarf galaxy, we find that our object would be classified as a starburst dwarf galaxy  \citep{lee:11,mcquinn:15B}.

Moreover, \cite{marasco:22} analysed a starburst dwarf galaxy sample from the DWarf galaxies Archival Local survey for Interstellar medium investigatioN (Dwalin; Cresci et al. in prep.), and showed the SFR as a function of the stellar mass of these galaxies. The SFR and stellar mass computed for our RPS-dwarf galaxy falls very close to two starburst dwarf galaxies: Tol65 and UM461.

\begin{figure}
    \centering
    {\includegraphics[width=0.46\textwidth]{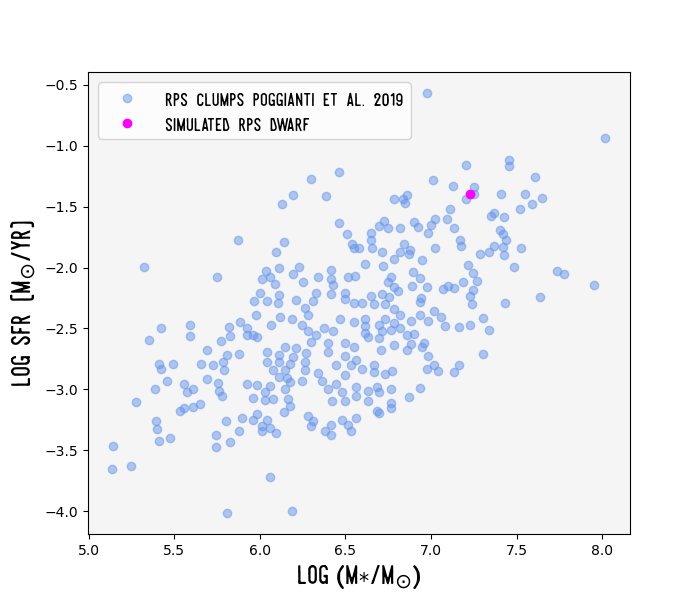} }\\
     {\includegraphics[width=0.46\textwidth]{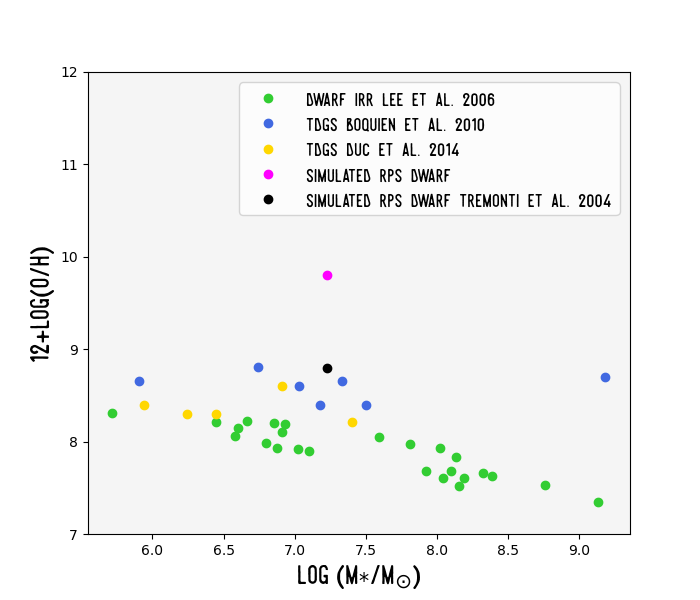}}
    \caption{The top panel shows the SFR as a function of stellar mass. We plot the star formation clumps found by \cite{poggianti:19} (blue dots) and compared those with the value of our RPS dwarf galaxy (magenta dot).
    The bottom panel shows the oxygen abundance as a function of stellar mass of all TNG-50 subhalos at z=0.1 (magenta dots), irregular galaxies from \cite{lee:06}(green dots), TDGs sample (blue dots) from \cite{boquien:10}, TDGs sample (yellow dots) from \cite{duc:14}. We compare the latter values with the oxygen abundance of our RPS dwarf galaxy from the TNG-50 data (magenta dot), and from scaling the \cite{tremonti:04}'s relation (black dot).
    }
    \label{Fig6}
\end{figure}

On the other hand, it is important to note that \cite{poggianti:19} studied star forming regions within the tails of jellyfish galaxies, finding over 500 of them. Their SFR range from $\sim 0.007$ to $1$ M$_{\odot}$/yr, and their stellar masses are in the range $M_{*} = 10^5 - 3\times10^7$ M$_{\odot}$, and sizes between $\sim100$ and $800$ pc. In Figure \ref{Fig6} we show the star forming regions reported by \cite{poggianti:19} (blue dots), and over-plotted the value of our RPS-dwarf galaxy (magenta dot). It is important to note that our RPS-dwarf galaxy is located where the RPS clumps have the highest values of SFR, and its mass is consistent with the jellyfish clump's distribution. 

Since our RPS-dwarf galaxy is being formed from already enriched material, we expect its metallicity to be higher than the metallicity of ``standard'' dwarf galaxies with a similar stellar mass. As we mentioned before, a better comparison in metallicity would be with TDGs, since they are also formed from recycled material previously metal-enriched. \cite{recchi:15}, compared the gas oxygen abundances of a sample of gas-rich dwarf galaxies \citep{lee:06} and a sample of TDGs \citep{duc:14,boquien:10}. They found that the oxygen abundance ($12+log(O/H)$) for the TDGs sample is $\sim 8-9$. Moreover, \cite{sweet:14} identified TDGs with $12+log(O/H)>8.6$. 

In the bottom panel of Figure \ref{Fig6} we compare the oxygen abundances of the gas-rich dwarf galaxies reported by \cite{lee:06} (green dots), and the TDGs reported by \cite{boquien:10} (blue dots), and \cite{duc:14} (yellow dots). 
We computed the oxygen abundance for our RPS-dwarf galaxy, obtaining  $12+log(O/H)=9.5$ (magenta dot in the bottom panel of Figure \ref{Fig6}), somewhat higher than those reported for star-forming ``standard'' dwarfs and TDGs. It is important to note that for the progenitor galaxy (ID 19447) $12+log(O/H)=10.15$, from such enriched gas our RPS-dwarf galaxy is formed, explaining its high value of oxygen abundance. Yet this value of oxygen abundance for the progenitor galaxy is high when compared to star-forming galaxies. In particular, using the mass-metallicity relation of \cite{tremonti:04}, inferred from 53 thousand star-forming galaxies at $z\sim0.1$, for the progenitor galaxy (subhalo with ID 119447) whose stellar mass is $8.65\times10^{10}$ M$_{\odot}$ we obtain $12+log(O/H) =9.1$, one dex lower than the value from TNG-50. Moreover, if we take the luminosity-metallicity relation of \cite{tremonti:04} one computes a value of $12+log(O/H) =9.2$. If the discrepancy of one dex is applied to our RPS-dwarf galaxy, we obtain a value of $12+log(O/H) =8.5$ (black dot in the bottom panel of Figure \ref{Fig6}), which matches the mass-metallicity region of TDGs in the bottom panel of Figure \ref{Fig6}.

We followed our detected RPS-dwarf galaxy forward in time (from z=0.1 to z=0). We found that it survives slightly more than $1$ Gyr. After that, our RPS dwarf falls back to the main galaxy and starts to be tidally disrupted by it (at z=0). Despite its fate, if we take the definition of \cite{bournaud:10}, according to whom a TDG is defined as a long lived object if its age is at least $1$ Gyr, then our RPS dwarf galaxy would be classified as a long lived object.

\section{Conclusions}
\label{sec:conclusions}
Detail observations of SF regions in the tentacles of jellyfish galaxies \citep{moretti:18,moretti:20,poggianti:19,giunchi:23} have been able to identified individual SF regions along their tentacles. Such SF regions are observed to have stellar masses up to $3\times10^7$ M$_{\odot}$, and have sizes up to $800$ pc. 

If a SF region (triggered by RPS) within the tentacles of jellyfish galaxies is self-bound (with the latter physical parameters of mass and size), it could give rise to a new type of intracluster dwarf galaxy; a \textit{stripped purely baryonic dwarf galaxy} \citep{kapferer:08,poggianti:19}. More recently, \cite{werle:24} have studied the stellar populations of the star forming complexes in the tails of strongly RP- stripped galaxies, finding their characteristics to be consistent with those of the population of dwarf galaxies in clusters. We use the state of the art cosmological-magneto-hydrodynamic simulation TNG-50, to look if such dwarf galaxies exists in simulations, and if they could be born solely as a consequence of RPS.

We analyzed a population of jellyfish galaxies in the cosmological simulation TNG-50, to look for most massive stellar over-densities within their long gas tentacles. We found a particularly large star forming region at the tip of a tentacle of the jellyfish galaxy with ID-119447. 

The stellar component of our RPS dwarf galaxy is made of 253 stellar particles, and 2036 gas particles. The latter number of particles guarantees that our RPS dwarf galaxy is well resolved 
(\cite{jochi:21} classified the lower stellar mass limit of dwarf galaxies in TNG-50 as those with 120 stellar particles).

We have to point out, that the resolution of TNG-50, prevent us to study smaller ($<0.5$ ckpc/h) star forming objects in the tail of jellyfish galaxies (as independent self-bound objects), and the ones (well resolved) detected in the simulation would be biased to be larger. Still, that was our goal: Finding dwarf-galaxy sized objects formed from the RPS gas in jellyfish galaxies.

We found that this region is self-bound, has a half-mass radius $r_{h,*}\approx1$ ckpc/h, and comprises a population of gas and stars ($M_{*,dwarf} = 1.7\times10^7$ M$_{\odot}$ and $M_{gas,dwarf} = 2\times10^8$ M$_{\odot}$). 

This dwarf-galaxy-like object has a concentration parameter $C=2.63$, it has a smooth stellar mass-distribution that can be approximated with a Plummer model. Its SFR has a value of $0.04$ M$_{\odot}$/yr, in agreement with the SFR values for star forming galaxies with similar stellar mass  \citep{mcquinn:15B,marasco:22}. It is metal rich according to TNG-50 ($12+\log(O/H)=9.8$), or ($12+\log(O/H)=8.5$) as derived from the metallicity-mass relation of \cite{tremonti:04} for its stellar mass, and it does not have a DM component.

The existence of such an object in the TNG-50 simulation, proves it is possible to form a dwarf galaxy similar in mass and size as those of standard dwarf galaxies, via RPS. In this work we imposed strict selection criteria to find the optimal galaxy from where a RPS dwarf galaxy could form (a SF region with an in-situ stellar population with M$_{*}>= 10^7$ M$_{\odot}$), but many more RPS dwarfs might be present in TNG-50, and even more low-mass SF regions associated to the tentacles of jellyfish galaxies. As an example, there is a region in the bottom left side of the top panel of Figure \ref{Fig2} where an elongated gas structure can be identified showing at its tip, a stellar over-density. It could be a second RPS dwarf formed in this galaxy. Therefore, in a following paper we will study the formation of RPS-dwarf galaxies in a wider range of radius and masses and try to ascertain how commonly this new type of dwarf galaxy can arise, as well as study the dynamical evolution (and fate) of RPS-dwarf galaxies. 

Using a state of the art cosmological simulation, we corroborate the galaxy formation scenario, which forms dwarf galaxies via RPS: \textit{ram pressure stripped dwarf galaxies}. These RPS dwarf galaxies are second-generation galaxies as they form from recycled (previously metal enriched) gas, and by construction do not have a DM halo. We find for the first time that these RPS dwarf galaxies could be as large as standard dwarf galaxies. Then, RPS dwarf galaxies would be a different class of dwarf galaxies, whose physical properties resemble those of standard dwarf galaxies.\\ 

\acknowledgments
We kindly thank the referee for the comments and suggestions which led to an improved version of this paper.
VL gratefully acknowledges support from the \mbox{CONACyT} Research Fellowship program. VL acknowledge support from PAPIIT-UNAM grant IN113522. VL kindly thanks E. K. Grebel, and F. J. Sánchez-Salcedo for very interesting and fruitful discussions about this paper. VL thanks specially A. Raga whom now rides the highest waves.
RS acknowledges support from the Fondecyt grant 1230441. JF acknowledges financial support from the UNAM- DGAPA-PAPIIT IN110723 grant, Mexico.

\end{document}